\documentstyle [referee,epsfig]{l-aa}

\begin{document}

\thesaurus{06(06.06.4; 06.15.1)}

\title{ Helioseismic tests of diffusion theory}

\author{G.~Fiorentini\inst{1,2} \and
	M.~Lissia\inst{3,4} \and
        B.~Ricci\inst{1,2} 
       }

\offprints{B. Ricci, ricci@fe.infn.it}

\institute{
 Istituto Nazionale di Fisica Nucleare, Sezione di Ferrara,
        via Paradiso 12, I-44100 Ferrara, Italy
\and  Dipartimento di Fisica dell'Universit\`a di Ferrara,
        via Paradiso 12, I-44100 Ferrara, Italy
\and  Istituto Nazionale di Fisica Nucleare, Sezione di Cagliari,
        Cittadella Universitaria, I-09042 Monserrato, Italy
\and  Dipartimento di Fisica dell'Universit\`a di Cagliari,
        Cittadella Universitaria, I-09042 Monserrato, Italy
  }

\date{Received..... / Accepted...... } 
  \maketitle

  \begin {abstract}
We present a quantitative estimate of the accuracy
of the calculated diffusion  coefficients, by comparing predictions of 
solar models with observational data provided by helioseismology.
By taking into account the major uncertainties in building solar
models we conclude that helioseismology confirms the diffusion efficiency
adopted in SSM calculations, to the $10\%$ level.
\end {abstract}  

\section{Introduction}
  \label{sec:intr}

\begin{figure}[htb]
\epsfig{file=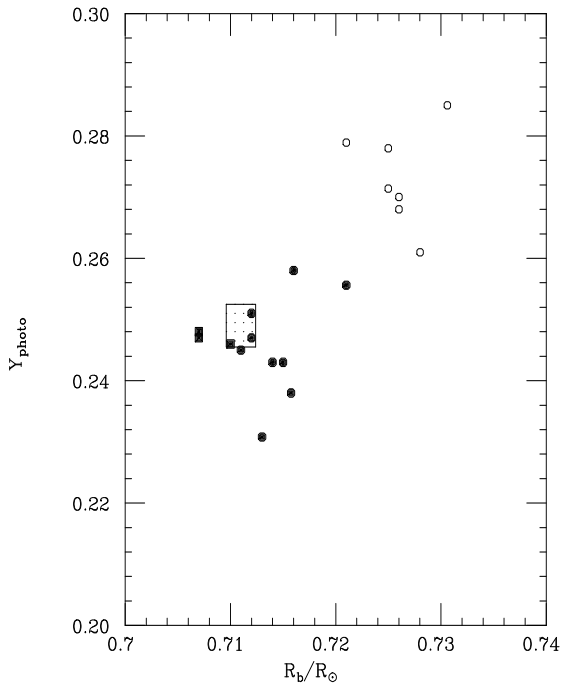,height=12cm,width=8cm}
\vspace{-2cm}
\caption[cc]
{The photospheric helium mass fraction $Y_{ph}$ and the depth
of the convective zone ($R_b$/$R_\odot$).
The dotted rectangle is the $1\sigma$ interval allowed by
helioseismology from Degl'Innocenti et al. (1997);
the open circles are the predicted values by solar models without diffusion, from
top to bottom correspond to Dar \& Shaviv (1996), 
Christensen-Dalsgaard  et al. (1993), Richard et al. (1996), 
Turck-Chi\`{e}ze \& Lopes (1993), Proffit (1994), Bahacall \& Pinsonneault (1995),
Ciacio et al. (1996).
The full squares are the predicted values by solar model with 
helium diffusion,  from top
to bottom  correspond to
Christensen-Dalsgaard  et al. (1993), Bahacall \& Pinsonneault (1992),
Proffit (1994).
The full circles are the predicted values 
by solar model with helium and heavy 
elements diffusion, from top to bottom correspond to  
Richard et al. (1996), Cox et al. (1989), Proffit (1994), 
Bahacall \& Pinsonneault (1995), Christensen-Dalsgaard  et al. (1996),
Turck-Chi\`{e}ze et al. (1998),
Bahacall  et al. (1998), Ciacio et al. (1996), Dar \& Shaviv (1996).
}
\label{figquadrato}
\end{figure}

In recent years inclusion of elemental
diffusion has been
an essential ingredient of stellar evolutionary codes for achieving
agreement between predicted and helioseismic values of properties of the
convective envelope, see e.g. Cox et al.(1989), Bahcall \& Pinsonneault (1995),
Bahcall et al. (1997), Ciacio et al. (1996), Degl'Innocenti et al. (1997),
Richard et al. (1996), Turck-Chi\`{e}ze et al. (1998),
and Fig. \ref{figquadrato}.
The success of solar models with diffusion, and the corresponding failures
of models  that neglect diffusion, suggest that the diffusion process has
been properly treated.
The goal of this paper is to present a quantitative estimate of the accuracy
of the calculated diffusion  coefficients, by comparing predictions of 
solar models with observational data provided by helioseismology.

Until recently, few stellar evolution codes included  diffusion.
Apart from practical difficulties, the omission of diffusion was 
justified since elemental diffusion occurs on a large time scale
(larger than $10^{13}$ yr. to diffuse a solar radius under solar
condition), which should imply that the effect on stellar structure are
small.
Nevertheless, the precision required for calculating  solar neutrino fluxes
and for matching the accuracy of helioseismic observations
is so great that  diffusion has to be taken into account, see e.g.
Bahcall \& Loeb (1990), Proffit (1994).

Most importantly, diffusion together with gravitational settling hide
below the convective envelope a significant fraction (about 10\%)
of the  initial  helium and heavy metals, and this effect
is crucial when comparing solar models with the observed 
properties of the convective envelope (e.g.the  present heavy element
abundance and helium fraction, this latter inferred by means of
helioseismology).

One of the most detailed calculations of diffusion coefficients
has been presented by Thoul et al. (1994), hereafter TBL.
It is based upon exact numerical solutions of the fundamental equations
for element diffusion and heat transfer discussed in Burgers (1969).
Good agreement exists between the diffusion rates computed by Thoul et al.
 and the results obtained with a  very different treatment of 
Michaud \& Proffit (1993).
The accuracy of the TBL coefficients is estimated to be 
about 15\% by comparison with the results of other authors, see 
Bhacall and Pinsonneault (1995)
and refs. therein.

However, some warning is needed:  the separation of elements
by diffusion might be inhibited by the presence of mixing
in the solar interior, see Schatzman (1969), Lebreton \& Maeder (1987),
Pinsonneault et al. (1989), Pinsonneault et al. (1990), Chaboyer et al. (1992),
Chaboyer et al. (1995).
We also remind the anomalous lithium depletion of the sun
with respect to meteorites, signalling some deficiency
of present standard solar models.

Clearly quantitative observational constraints on the
calculated diffusion coefficients would be welcome. The
purpose of this paper is to discuss the constraints posed 
by helioseismology.

As previously remarked  recent Standard Solar Models (SSMs)
which include diffusion, as well as
accurate equations of state and updated opacity
tables, are in good agreement with helioseismology. 
In essence we address the following question:
{\em which variation of the diffusion efficiency is allowed
without spoling the agreement with helioseismic data?}

In the next section we study how the predictions of solar models
are affected when the diffusion efficiency is changed and 
we  summarize the relevant helioseismic 
information.

In sect. \ref{sec3} we discuss the comparison of helioseismic data 
with the predictions
of solar models, when the diffusion efficiency is varied.

\section{The diffusion efficiency and solar models predictions}
\label{sec2}

For our standard solar model calculations we use 
the diffusion coefficients calculated following TBL.
For each element one has to consider
three different coefficients ($A_P, A_T,$ and $A_H$ following
the notation of TBL), each depending on the value of several
chemical and physical variables ($\rho, T, X, Z_i...$) in the solar
interior, so that in principle one has
really many parameters which could be varied for 
numerical studies.
All the diffusion coefficients, however, are approximately constant
along the solar profile apart from the innermost
core, see fig. 11 in TBL.

For simplicity, we shall introduce just one {\em diffusion efficiency 
parameter} $d$ and we shall study solar models obtained
by rescaling the TBL coefficients by the factor $d$, assumed
to be the same for all elements at any point in the solar
interior. 

By numerical experiments with our evolutionary code FRANEC
(Cieffi \& Straniero (1988))
 we have determined solar models (i.e.
stellar structures with luminosity $L_\odot= 3.844\cdot 10^{33}$erg/s,
radius $R_\odot=6.9599\cdot 10^{10}$cm and photospheric abundance 
$Z/X_{ph}=0.0247$ at age $t_\odot=4.57$ Gy for 
a mass $M_\odot= 1.989\cdot 10^{33}$ g) obtained for different values of $d$.

By means of helioseismology one can reconstruct accurately
three independent properties of the convective envelope:
its depth $R_b$, its density at the bottom $\rho_b$ and
the photospheric helium abundance $Y_{ph}$.
The resulting values together with estimated
$1\sigma$ uncertainties are shown in Table \ref{tabhelios}
together with SSM predictions from the model with
helium and heavy element diffusion of Bahcall and Pinsonneault (1995) 
hereafter BP95.

\begin{table}
\caption[errori]{
The solar observables $Q_i$ used in our analysis, their predicted
values $Q_{SSM i}$ from BP95, the helioseismic
determination $Q_{\odot i}$ with their $1\sigma$
error $\Delta Q_i$, from Degl'Innocenti et al. (1997)
 and the calculated 
coefficient $\alpha$, corresponding to Eq. \ref{eqQ}.
}
\begin{tabular}{lcccc}
\hline
$Q_i$ &  $Q_{SSM i}$  & $Q_{\odot i}$ & $\Delta Q_i$ & $\alpha_Q$\\
\hline
$Y_{ph}$& 0.247 &0.249 & 0.0035 & -0.091 \\
$R_b/R_\odot$&  0.712 & 0.711 & 0.0014 & -0.016\\
$\rho_b$ [g/cm$^3$] & 0.187 & 0.192 &0.0018 & 0.14 \\
\hline
\end{tabular}
\label{tabhelios}
\end{table}


By calculating these observables
for our solar models, we have studied their dependence
on $d$.
For each observables $Q$ we introduce a parametrization of the
form:
\begin{equation}
\label{eqQ}
Q(d)= Q_{SSM} d^{\alpha_Q}
\end{equation}
The coefficients $\alpha$ are collected in Table \ref{tabhelios}. 

The numerical results can be qualitatively explained 
by simple considerations. Consider as an example
the case of increased diffusion efficiency ($d>1$),
with respect to the SSM ($d=1$):\\
i) As a larger fraction of helium 
 is hidden below the convective
envelope, whereas  the initial helium abundance is essentially fixed
by the present luminosity, this implies a lower photospheric
helium than in the SSM;\\
ii)the metal abundance below the convective envelope
(for a fixed photospheric abundance) is increased
so that opacity increases and convection starts deeper and
consequently at higher density.
In addition from helioseismology one can reconstruct 
the sound speed profile of the solar interion 
with a  $1\sigma$ accuracy of about 0.15\% in the intermediate
region ($0.2< R/R_\odot< 0.6$).

This piece of information, however, is not particularly
illuminating for studying of diffusion. As already remarked
in Degl'Innocenti et al. (1997), solar models without diffusion
can have sound speed profile quite consistent with the 
helioseismic one.

\section{Diffusion efficiency and the properties of the convective
envelope}
\label{sec3}

We now determine the acceptable range of $d$ by requiring that
the three independent observables of the convective envelope,
$R_b$, $\rho_b$ and $Y_{ph}$, are predicted within their helioseismic ranges, 
by using Eqs.~(\ref{eqQ}) to determine the dependence of these properties
on $d$. 

We remind that
there are at least three major uncertainties in building standard
solar models that also have the potentiality of affecting the three
properties under investigation, and, therefore, that could
interfere with/hinder the effect of $d$: the astrophysical factor
$S_{pp}$, the solar opacity $\kappa$, the heavy element abundance
$\zeta = Z/X$. We shall add all these effects
one after the other, and determine a range of $d$ that
takes into account these uncertainties.

We start by defining a $\chi^2$ as:
\begin{equation}
\label{chi2_0}
\chi^2(d) = \sum_{i=1,3} 
\left ( \frac{Q_i(d)-Q_{\odot i}}{\Delta Q_i} \right )^{2} \, ,
\end{equation}
where the sum include the three independent observables of the
convective envelope, 
$Q_i (d)$ are computed by using Eqs.~(\ref{eqQ}), and $Q_{\odot i}$ are
the helioseismic values reported in Table \ref{tabhelios}
together with the $1\sigma$ errors $\Delta Q_i$.
The value $\chi^2(d=1)$ indicates how well the SSM
reproduces these  helioseismic properties. The first row of
Table~\ref{tabdelta} shows the good agreement between  BP95 
and helioseismology ($\chi^2$/dof = 8.61/3). 

If we use $d$ as free parameter (second row of Table~\ref{tabdelta}),
we find the following best fit value ($\chi^2$/dof = 2.92/2) and $1\sigma$
range:
\begin{equation}
\label{rangedeltanospp}
  d = (1.14 \pm 0.06)  \, ,
\end{equation}
i.e. helioseismology suggests a diffusion efficiency slightly 
larger than the SSM estimate.

\begin{table}
\caption[delta]{Deviations from standard diffusion allowed by helioseismic
measurements. The first five columns show whether the parameter is kept
fixed (F) at its SSM value or it is allowed to vary (V) as a free parameter.
The sixth
column shows the resulting $\chi^2$ per degree of freedom. The last two
columns show the best fit value for $d$ and its $1\sigma$ error. 
 }
\begin{tabular}{ccccccc}
\hline
 $\kappa$ & $\zeta$ & $S_{pp}$ & $d$ & $\chi^2$/dof &
           $d_{\mathrm{Best}} $ & $\Delta d $ \\
\hline
  F  &  F  &  F  & F  &  8.61 /3  &       &      \\
  F  &  F  &  F  & V  &  2.92 /2  &  1.14 & 0.06 \\
  F  &  F  &  V  & V  &  1.81 /2  &  1.07 & 0.09 \\
  F  &  V  &  V  & V  &  0.98 /2  &  1.05 & 0.09 \\
  V  &  V  &  V  & V  &  0.75 /2  &  1.08 & 0.11 \\
\hline
\end{tabular}
\label{tabdelta}
\end{table}

\subsubsection{Uncertainties on $S_{pp}$}
A conservative estimate of the uncertainty is provided by the range of
the published results (NACRE coll. (1998)), whereas
a $1\sigma$ estimate has been provided in
Kamionkowski \& Bahcall (1994);
we shall use
$\Delta S_{pp}/S_{pp}^{\mathrm{SSM}} = 0.05/3$ at $1\sigma$
(5\% is the ``$3\sigma$ error'' estimate).
The dependence of $Q_i$ on $S_{pp}$ has been determined numerically in
Degl'Innocenti et al. (1998). By redefining a suitable $\chi^2(d,S_{pp})$:
\begin{eqnarray}
\label{chi2spp}
\chi^2(d, S_{pp}) &=& \sum_i
\left ( \frac{Q_i(d,S_{pp}) - 
                Q_{\odot i}}{\Delta Q_i} \right )^{2} \,  \nonumber \\
&+&\left ( \frac{S_{pp}- S_{pp,\mathrm{SSM}}}{\Delta S_{pp}} \right )^{2}\, ,
\end{eqnarray}
we find now:
\begin{equation}
\label{rangedeltaspp}
  d = (1.07 \pm 0.09) \, , 
\end{equation}

\subsubsection{Uncertainties on $\kappa$ and $\zeta$}
The heavy element abundance $\zeta$ and the solar opacity $\kappa$
are known with a conservative accuracy of about
$ 10$\%, see Bahcall \& Pinsonneault (1992), V. Castellani et al. (1997).
 Therefore,
our $1\sigma$ relative error estimate will be $ 0.1/3$.
The dependence\thanks{We remark that we are considering a constant rescaling
of opacity along the solar profile.} of $Q_i$ on $\kappa$ and $\zeta$ 
has been determined numerically in Degl'Innocenti et al. (1998). In this case,
the relevant $\chi^2$ is:
\begin{eqnarray}
\label{chi2sppzk}
\chi^2(d, S_{pp},\zeta,\kappa) &=& \sum_i
\left ( \frac{Q_i(d,S_{pp},\zeta,\kappa) - 
                Q_{\odot i}}{\Delta Q_i} \right )^{2}  \nonumber \\
 &+&
\left ( \frac{S_{pp}- S_{pp,\mathrm{SSM}}}{\Delta S_{pp}} \right )^{2} \, + \,
\left ( \frac{\kappa-\kappa_{SSM}}{\Delta \kappa} \right )^{2}  \nonumber \\
&+& 
\left ( \frac{\zeta -\zeta_{SSM}}{\Delta \zeta} \right ) ^{2} \, .
\end{eqnarray}
We find a small change of the best fit value and, of course, an
increase of the $1\sigma$ range of $d$:
\begin{equation}
\label{rangedeltasppzd}
  d= (1.08 \pm 0.11) \, .
\end{equation}

\subsubsection{Solar model ``theoretical uncertainties''}
At last we try to estimate how much our results could depend on having used
BP95 as reference standard model. 
To this end, we consider one of the standard solar models (models that
include all the state-of-the-art solar physics), whose helioseismic properties
differ the most from BP95 and, consequently, fit less well the
experimental data.
We repeated the above-described analysis by using FR97 (Ciacio et al. (1996)) as
standard solar model. When all parameters are varied, the $1\sigma$ range 
and best fit value, {\em cf}. Eq.~(\ref{rangedeltasppzd}), become:
\begin{equation}
\label{rangedeltaf97}
  d = (1.13 \pm 0.12)\, ,
\end{equation}
in substantial agreement with Eq.~(\ref{rangedeltasppzd}).\\

In conclusione, helioseismology confirms the diffusion efficiency
adopted in SSM calculations, to the $10\%$ level ($1\sigma$ C.L.).

\begin{acknowledgements}
We thank Thoul for providing us with the exportable
subroutine.
\end{acknowledgements}


\begin{thebibliography}{}  


\bibitem[1990]{BL90}
Bahcall, J.N., and Loeb, A., 1990,  ApJ 360, 267.

\bibitem[1992]{BP92}
Bahcall, J.N and Pinsonneault, M.H., 1992,  
Rev. Mod. Phys.  64, 885.

\bibitem[1995]{BP95}
  Bahcall, J.N and Pinsonneault, M.H., 1995,
  Rev. Mod. Phys. 67,  781.

 \bibitem[1997]{BPBCD97}
Bahcall, J.N.,  Pinsonneault, M.H., Basu, S., and Christensen-Dalsgaard, J.,
1997,  Phys. Rev. Lett. 78, 171.


\bibitem[1998b]{BP98}
 Bahcall,  J.N.,  Basu , S., and  Pinsonneault, M.H., 1998, astro-ph/9805135,
to appear  on Phys. Lett. B.


\bibitem[1969]{B69}
 Burgers, J.M., 1969, Flow Equations for Composite Gases, Academic Press,
New York.


\bibitem[1997b]{Report}
Castellani,  V., Degl'Innocenti, S.,  Fiorentini, G.,  Lissia, M., and Ricci,
B., 1997, Phys. Rep. 281,  309.


\bibitem[1995]{CDP95}
 Chaboyer,  B.,  Demarque, P.,  and  Pinsonneault, M.H., 1985,
 ApJ  441, 865.


\bibitem[1992]{CVZ92}
Chaboyer, B.,  Vauclair, S.,  and  Zahn,  J.P.,  1992, A\&A 
255, 191.

\bibitem[1993]b{CD}
  Christensen-Dalsgaard,  J.,  Proffitt,   C.R., and  Thompson, M.J., 1993,  
ApJ 403,  L75;


\bibitem[1996c]{JCD}
Christensen-Dalsgaard, J.,  et al.,   1996,  Science 272, 1286.
   


\bibitem[1996]{Ciacio}
Ciacio, F.,  Degl'Innocenti, S.,  and Ricci, B. , 1996,
A\&AS 123, 44

\bibitem[1988]{CS88}
Cieffi,  M., and  Straniero,  O., 1989, ApJS 71, 47.


\bibitem[1989]{C89}
Cox, A.N. ,  Guzic , J.A. and  Kidman,  R.B.,  1989,
 ApJ 342 , 1187.

\bibitem[1996]{DS96}
Dar, A.,  and Shaviv, G., 1996,  ApJ  468,   933.


 \bibitem[1997]{eliosnoi}
 Degl'Innocenti,   S., Dziembowski, W.A., Fiorentini, G., and
 Ricci,    B., 1997 Astroparticle. Phys. 7, 77.


  \bibitem[1998]{eliospp}
Degl'Innocenti,  S.,  Fiorentini,  G., and  Ricci, B., 1998,
   Phys. Lett. B 416,  365.


  \bibitem[1994]{KB94}
Kamionkowski,   M., and Bahcall, J.N., 1994,  ApJ  420, 884.


\bibitem[1987]{LM87}
 Lebreton, Y.,  and  Maeder, A., 1987,  A\&A  175,  99.


\bibitem[1993]{MP93}
 Michaud, G., and Proffit, C.R. , 1993 in `Inside the Stars, IAU Col. 137,
A. Baglin and W.W. Weiss eds., PASP, San Francisco.



  \bibitem[1998]{NACRE}
 NACRE collaboration, 1998, private comunication, to appear
  within the publication plane of the collaboration.


\bibitem[1989]{P89}
 Pinsonneault , M.H. et al., 1989,  ApJ  338, 424,

\bibitem[1990]{PKD90}
 Pinsonneault, M.H.,  Kawaler, S.D.,  and  Demarque, P., 1990, ApJS, 
74,  501.

\bibitem[1994]{P94}
 Proffit,  C.R., 1994, ApJ 425, 849.


 \bibitem[1996]{RCVD96}
 Richard,  O.,  Vauclair, S., Charbonnel,  C.,  and Dziembowski, W.A., 1996,
  A\&A 312, 1000.


\bibitem[1969]{S69}
Schatzman,  E., 1969, A\&A  3, 331.

\bibitem[1994]{TBL94}
Thoul,  A.A., Bahcall, J.N.,  and Loeb ,A., 1994, ApJ 421, 828.


\bibitem[1993]{TCL}
Turck-Chi\`{e}ze, S.,  and Lopes,  I.,  1993, ApJ  408, 347.


\bibitem[1998]{TC98}
 Turck-Chi\`{e}ze, S. et al., 1998, astro-ph/9806272.

\end{thebibliography}
\end{document}